\documentclass[prl,amsmath,amssymb,aps,twocolumn]{revtex4-1}

\usepackage{subcaption}
\usepackage[dvipsnames]{xcolor}
\usepackage{graphicx}
\usepackage[colorlinks=true,citecolor=blue]{hyperref}
\usepackage{soul}
\setstcolor{Green}
\usepackage{cancel}
\DeclareMathAlphabet{\mathcalligra}{T1}{calligra}{m}{n}
\usepackage[utf8]{inputenc}

\newcommand{\subR}{\textrm{\tiny{R}}}
\newcommand{\subE}{\textrm{\tiny{E}}}

\newcommand{\be}{\begin{equation} }
	\newcommand{\ee}{\end{equation}}
\newcommand{\bes}{\begin{equation*} }
	\newcommand{\ees}{\end{equation*}}
\newcommand{\bea}{\begin{eqnarray} }
	\newcommand{\eea}{\end{eqnarray}}
\newcommand{\beas}{\begin{eqnarray*} }
	\newcommand{\eeas}{\end{eqnarray*}}
\newcommand{\ba}{\begin{align} }
	\newcommand{\ea}{\end{align} }
\newcommand{\bas}{\begin{align*} }
	\newcommand{\eas}{\end{align*} }

\usepackage{caption}

\begin{document}
	
	\title{A mode-sum prescription for the renormalized stress energy tensor on black hole spacetimes}
	
	\author{Peter Taylor}
	\email{peter.taylor@dcu.ie}
	\affiliation{
		Center for Astrophysics and Relativity, School of Mathematical Sciences, Dublin City University, Glasnevin, Dublin 9, Ireland}
	
	\author{Cormac Breen}
	\email{cormac.breen@tudublin.ie}
	\affiliation{School of Mathematical Sciences, Technological University Dublin, Grangegorman, Dublin 7, Ireland}

	\author{Adrian Ottewill}
	\email{adrian.ottewill@ucd.ie}
	\affiliation{School of Mathematics and Statistics, University College Dublin, Belfield, Dublin 4, Ireland}

	\date{\today}
	\begin{abstract}
		In this paper, we describe an extremely efficient method for computing the renormalized stress-energy tensor of a quantum scalar field in spherically-symmetric black hole spacetimes. The method applies to a scalar field with arbitrary field parameters. We demonstrate the utility of the method by computing the renormalized stress-energy tensor for a scalar field  in the Schwarzschild black hole spacetime, applying our results to discuss the null energy condition and the semi-classical backreaction.	
	\end{abstract}
	\maketitle
	\section{Introduction}
	The semi-classical approximation to quantum gravity has by now a long and fruitful history. In particular, Parker's discovery of particle production in an expanding universe \cite{Parker1968} and Hawking's discovery that black holes radiate \cite{Hawking:1975} have had a profound impact on theoretical physics. Formally, this approximation involves the propagation and backreaction of quantized fields on a classical spacetime geometry described by the semi-classical equations
	\begin{align}
		G_{ab}-\Lambda\,g_{ab}=8\pi\,\langle\hat{T}_{ab}\rangle,
	\end{align}
	where $g_{ab}$ is the metric of spacetime, $G_{ab}$ is the Einstein tensor, $\Lambda$ is the cosmological constant and $\langle \hat{T}_{ab}\rangle$ is the (unregularized) expectation of the stress-energy tensor of a quantum field in some quantum state. An immediate difficulty arises in this framework, $\langle \hat{T}_{ab}\rangle$ diverges everywhere and hence a regularization prescription is required. Formally, this prescription is well understood through, for example, point-separation \cite{Christensen:1976}. This amounts to considering instead a bitensor $\langle\hat{T}_{ab}(x,x')\rangle$ which is evaluated at two distinct spacetime points, then we can isolate and subtract the divergent terms as $x'\to x$. The divergent terms are geometrical in nature, depending only on the metric and its derivatives. The semi-classical equations then become
	\begin{align}
		G_{ab}-\Lambda\,g_{ab}+\alpha H^{(1)}_{ab}+\beta\,H^{(2)}_{ab}=8\pi\,\langle\hat{T}_{ab}\rangle_{\subR},
	\end{align}
	where the right-hand side is now the renormalized stress-energy tensor (RSET) and $H^{(1)}_{ab}$, $H^{(2)}_{ab}$ are geometrical terms that are quadratic in curvature. This form of the semi-classical equations illuminates the prescription for making sense of the theory, the stress-tensor is regularized by point-separation, then given physical meaning through renormalizing the constants $\Lambda$, $\alpha$ and $\beta$.
	
	While the point-separation prescription offers a formal resolution for regularizing the semi-classical theory, there remains a stubborn technical challenge with its implementation. To elucidate this challenge,  let us assume the spacetime is static and that the quantum field is in the Hartle-Hawking state \cite{HartleHawking:1976, Jacobson:1994}. In this case, we can work on the Euclideanized version of the spacetime. We note that the point-split stress-energy tensor for the Hartle-Hawking state is related to a differential operator acting on the (Euclidean) Green function.  In order to renormalize, we subtract from the  Green function a two-point distribution known as a Hadamard parametrix \cite{FullingWald:1978} before taking the limit $x'\to x$. 
	The parametrix is locally constructed in such a way that subtracting from the  Green function results in a smooth two-point function in this limit.  
	In 4-dimensional space-time,  we may take the Hadamard parametrix to be
	\begin{align*}
		K(x,x') = \frac{1}{8\pi^2} \left( \frac{\Delta^{1/2}(x,x')}{\sigma(x,x')} + V(x,x') \ln\left( \frac{2\sigma(x,x')}{\ell^{2}}\right) \right) 
	\end{align*}
	where $2\sigma(x,x')$ is the square of the geodesic distance between two nearby spacetime points $x$ and $x'$, $\Delta(x,x')$ is the Van Vleck-Morrette determinant and $V(x,x')$ is a symmetrical geometrical biscalar which encodes how waves scatter off the geometry of the spacetime. The parameter $\ell$ is an arbitrary lenghtscale required to make the argument of the logarithm dimensionless; it is essentially the well-known renormalization ambiguity (see, for example, Ref.~\cite{waldBook:1994}). The important point is that $K(x,x')$ is locally constructed from the spacetime geometry through the metric and its derivatives
	
	On the other hand, the   Green function $G(x,x')$ is not locally constructed but depends on global properties of the spacetime; we typically write $G(x,x')=K(x,x') + W(x,x')$ where $W(x,x')$ is a non-geometrical (symmetric) biscalar that encapsulates the global properties.  $G(x,x')$ is typically expressed as a mode-sum representation where, for example, global information such as the quantum state is encoded in boundary conditions on the individual modes. The singular behaviour in the limit where spacetime points are taken together is manifest as the non-convergence of this mode-sum. The technical challenge then is expressing the local Hadamard parametrix as a mode sum that can be subtracted from the Euclidean Green function mode by mode, producing a sum which converges in the limit where spacetime points coincide.
	
	The first resolution to this technical problem for a quantum field on a black hole spacetime was given in a seminal paper by Candelas and Howard \cite{CandelasHoward:1984}. Other approaches with a degree of novelty relied heavily on the essential ideas in the Candelas-Howard prescription \cite{AHS1995, TaylorOttewill2010, BreenOttewill2012}.  Notwithstanding the ingenuity of the method, it is cumbersome and inefficient to implement. In recent years, there have been two new developments in this area. The first, known as the ``pragmatic mode-sum prescription'' was developed by Levi and Ori \cite{LeviOri:2015, LeviOri:2016}. The method has proven indeed to be pragmatic, both in its efficiency and its broader applicability. Of particular note is the application of the prescription to compute the RSET for a scalar field on a Kerr black hole \cite{LeviOriKerr:2017}, a longstanding problem in the QFTCS community. Second is a method developed by the authors of this article known as the ``extended coordinate method'' \cite{taylorbreen:2016,taylorbreen:2017}. While this method has thus far only been developed for computing the simpler vacuum polarization in static black hole spacetimes, it is extremely efficient and applicable to arbitrary field parameters and arbitrary spacetime dimensions. 
	
	In this paper, we present the extended-coordinate method for the calculation of the RSET for an arbitrary scalar field in four dimensions (though the extension of this method to higher dimensions is straightforward). As an example of our method, we present results for the RSET of a scalar field in the Schwarzschild black hole spacetime with various values of field mass and coupling constant. 
	As applications of our results, we examine (a)~dependence of the RSET components on the coupling (b)~the semi-classical backreaction and (c)~the null energy conditions on the photon sphere.
	
	\section{Renormalizing the Green Function}
	We consider a quantum scalar field on a static, spherically symmetric black hole spacetime. Since we will assume the field is in the Hartle-Hawking quantum state, it is appropriate and convenient to work with the Euclideanized line element
	\begin{align}
		\label{eq:metric}
		ds^{2}=f(r)d\tau^{2}+dr^{2}/f(r)+r^{2}( d\theta^2 + \sin^2 \theta d \phi^2).	\end{align}
	It can be shown that the Euclidean metric would possess a conical singularity on the horizon unless we enforce the periodicity $\tau=\tau+2\pi/\kappa$ where $\kappa$ is the surface gravity. Imposing this periodicity discretizes the frequency spectrum of the field modes which now satisfy an elliptic wave equation
	\begin{align}
		(\Box_{\subE}-\mu^2-\xi\,R)\phi=0,
	\end{align}
	where $\Box_{\subE}$ is the d'Alembertian operator with respect to the Euclidean metric, $\mu$ is the field mass, $R$ is the Ricci curvature scalar of the background spacetime and $\xi$ is the coupling strength between the field and the background geometry. The corresponding Euclidean Green function has the following mode-sum representation
	(with $r=r'$ for simplicity): 
	\begin{align}
		\label{eq:Gmodesum}
		&G(x,x')=\frac{1}{8 \pi^2}\sum_{l=0}^{\infty}(2l+1)P_l(\cos\gamma)\sum_{n=-\infty}^{\infty}e^{in\kappa\Delta\tau}g_{nl}(r),
	\end{align}
	where $\Delta x\equiv x'-x\sim\mathcal{O}(\epsilon)$ is the coordinate separation, $\gamma$ is the geodesic distance on the 2-sphere and $\kappa$ is the surface gravity of the black hole horizon. We have also taken $P_{l}(z)$ to be the Legendre polynomial of the first kind and $g_{nl}=\kappa\,N_{nl}p_{nl}(r)\,q_{nl}(r)$ is the one-dimensional radial Green function evaluated at the same spacetime point $r$. The radial modes $p_{nl}(r)$, $q_{nl}(r)$ are solutions of the homogeneous radial equation:
	\begin{align*}
		\bigg[\frac{d}{dr}\Big(r^2f(r)\frac{d}{dr}\Big)-r^{2}\left(\frac{n^{2}\kappa^{2}}{f(r)}+(m^{2}+\xi\,R)\right)\nonumber\\
		-l(l+1)\bigg]Y_{nl}(r)=0,
	\end{align*}
	where $p_{nl}(r)$ and $q_{nl}(r)$ are regular on the horizon and the outer boundary (usually spatial infinity), respectively. The normalization constant is given by
	\begin{align*}
		N_{nl}=-r^{2}f(r)\,W\{p_{nl}(r),q_{nl}(r)\},
	\end{align*}
	where $W\{p,q\}$ denotes the Wronskian of the two solutions.
	
	In the coincidence limit $\Delta x\to 0$ (i.e. $\gamma\to 0$ and $\Delta\tau\to 0$), the mode-sum (\ref{eq:Gmodesum}) diverges. To renormalize this mode sum, we must find a way to express the locally-constructed Hadamard parametrix $K(x,x')$ as a mode sum and subtract mode-by-mode. In \cite{taylorbreen:2016,taylorbreen:2017} a mode sum expression for the Hadamard parametrix was derived by first introducing the so-called extended coordinates:
	\begin{align*}
		\varpi^{2}=\frac{2}{\kappa^{2}}(1-\cos \kappa\Delta\tau),\quad s^{2}=f(r)\,\varpi^{2}+2 r^{2}(1-\cos\gamma).
	\end{align*}
	For simplicity, the separation in the radial direction, $\Delta r$, is set to zero but it is important to the development that the separation in the other directions is maintained. Expressing the Hadamard parametrix in terms of these extended coordinates permits  
	its decomposition in terms of Fourier frequency modes and multipole moments where, remarkably, the coefficients in this decomposition are expressible in closed form for any static spherically-symmetric spacetime in arbitrary dimensions. The details are rather complicated and the expressions lengthy so we relegate them to the Appendix, giving only a schematic representation below. In four dimensions, the result is
	\begin{align}
		\label{eq:HadamardExp}
		K(x,x')&=\frac{1}{8\pi^{2}}\sum_{l=0}^{\infty}(2l+1)P_{l}(\cos\gamma)\sum_{n=-\infty}^{\infty}e^{in\kappa\Delta\tau}k_{nl}(r)\nonumber\\
		&+\frac{1}{8 \pi^2}\Big\{\mathcal{D}^{(-)}_{11}(r) + \left(\mathcal{T}^{(p)}_{10}+	\mathcal{D}^{(-)}_{22}(r)\right) s^2\nonumber\\
		&+\left(\mathcal{T}^{(p)}_{11}+\mathcal{D}^{(-)}_{21}(r)\right)\varpi^2 \Big\}+ O(\epsilon^{2m} \log \epsilon),
	\end{align}
	where the mode-sum regularization parameters are contained in $k_{nl}(r)$ which we further express as
	\begin{align}
		\label{eq:RegParam}
		k_{nl}(r)=& \sum_{i=0}^{m}\sum_{j=0}^{i}\mathcal{D}^{(+)}_{ij}(r)\Psi^{(+)}_{nl}(i,j|r)\nonumber\\
		&+\sum_{i=0}^{m-1}\sum_{j=0}^{i}\mathcal{T}^{(l)}_{ij} \chi_{nl}(i,j|r)\nonumber\\
		&+\sum_{i=1}^{m-1}\sum_{j=0}^{i-1}\mathcal\mathcal{T}^{(r)}_{ij} \Psi^{(+)}_{nl}(i+1,j|r).
	\end{align}
	Here $m$ denotes the order of the expansion. The coefficients $\mathcal{D}^{(\pm)}_{ij}(r)$ arise in the expansion of the direct part $\Delta^{1/2}/\sigma$ of the Hadamard parametrix when expanded in extended coordinates $s$ and $\varpi$, while the $\mathcal{T}^{(l)}_{ij}(r)$, $\mathcal{T}^{(p)}_{ij}(r)$ and $\mathcal{T}^{(r)}_{ij}(r)$ arise in the expansion of the tail $V\,\log(2\sigma)$. When the tail part of the Hadamard parametrix is expanded in $s$ and $\varpi$, we obtain terms that are logarithmic in $s$, polynomial in $s^{2}$ and $\varpi^{2}$ and rational in $s^{2}$ and $\varpi^{2}$. It is the coefficients of these terms that we have labelled $\mathcal{T}^{(l)}_{ij}(r)$, $\mathcal{T}^{(p)}_{ij}(r)$ and $\mathcal{T}^{(r)}_{ij}(r)$, respectively. The terms $\Psi_{nl}^{(+)}(i,j|r)$ and $\chi_{nl}(i,j|r)$ are the so-called regularization parameters that arise in  expressing $K(x,x')$ as a mode-sum. In particular, the $\Psi_{nl}^{(+)}(i,j|r)$ are obtained by representing terms of the form $\varpi^{2i+2j}/s^{2j+2}$ in a multipole and Fourier decomposition
	\begin{align}
		\frac{\varpi^{2i+2j}}{s^{2j+2}}=\sum_{n=-\infty}^{\infty}e^{in\kappa\Delta\tau}\sum_{l=0}^{\infty}(2l+1)P_{l}(\cos\gamma)\Psi_{nl}^{(+)}(i,j|r)  
	\end{align}
	and then using the completeness relations to invert these to obtain closed-form representations of $\Psi_{nl}^{(+)}(i,j|r)$. The particular form is given in the Appendix. Similarly, the $\chi_{nl}(i,j|r)$ regularization parameters are obtained by inverting
	\begin{align}
		&s^{2i-2j}\varpi^{2j}\log\left(s^{2}/\ell^{2}\right)\nonumber\\
		&=\sum_{n=-\infty}^{\infty}e^{in\kappa\Delta\tau}\sum_{l=0}^{\infty}(2l+1)P_{l}(\cos\gamma)\chi_{nl}(i,j|r),
	\end{align}
	where $\ell$ on the left-hand side is the arbitrary lengthscale. Again, explicit expressions for $\chi_{nl}(i,j|r)$ are found in the Appendix. 
	
	The important point is that the Hadamard parametrix expressed in the form (\ref{eq:HadamardExp}) can be subtracted from the corresponding mode-sum expression for the Green function (\ref{eq:Gmodesum}) yielding a mode sum expression for the full, renormalized Green function that is convergent in the coincidence limit. Moreover the speed of convergence of this mode sum can be accelerated by subtracting a higher-order expansion of the singular field. It is worth noting that the error term in Eq. (\ref{eq:HadamardExp}) ignores terms that are polynomial in $s^2$ and $w^2$, as the mode sum decomposition of these terms does not aid in the convergence of the mode sums, see \cite{taylorbreen:2017} for further details.
	
	As a concrete example, this prescription for the vacuum polarization in the Hartle-Hawking state yields
	\begin{align}
		\label{eq:VP}
		\langle \hat{\phi}^{2}\rangle_{\subR}&\equiv \left[W(x,x')\right]\equiv w(r)\nonumber\\
		&=\frac{1}{8\pi^{2}}\sum_{l=0}^{\infty}\sum_{n=-\infty}^{\infty}(2l+1)\mathsf{g}_{nl}-\frac{\mathcal{D}_{11}^{(-)}(r)}{8\pi^{2}}.
	\end{align}
	where, as above,  $W(x,x')\equiv G(x,x')-K(x,x')$ and we have adopted the notation
	\begin{align}
		\mathsf{g}_{nl}(r)\equiv g_{nl}(r)-k_{nl}(r)
	\end{align}
	for the 
	renormalized modes. Here and throughout, we have adopted square brackets $[\cdot]$ to denote the coincidence limit $x'\to x$.
	
	The modes $\mathsf{g}_{nl}(r)$
	converge like $O(l^{-2m-3})$ for large $\ell$, fixed $n$ and $O(n^{-2m-3})$ for large $n$, fixed $\ell$ so that the sum in (\ref{eq:VP}) converges very rapidly for sufficiently high expansion order $m$. In practice, provided $m$ is chosen appropriately, the sum in (\ref{eq:VP}) can be computed to very high accuracy by summing only a handful of $l$ and $n$ modes.
	
	\section{Computing the RSET}
	The calculation of the RSET has the potential to be much trickier as the components of the RSET involve derivatives with respect to $r$ and $r'$ while our expansion above assumed that radial points are taken together.

	The Euclidean Green function describing our state may be expressed as $G(x,x') =K(x,x')+ W(x,x')$ where $W(x,x')$ is regular near coincidence and symmetric in $x$ and $x'$.
	Correspondingly $W(x,x')$ has a covariant Taylor series for $x'$ near $x$ of the form~\cite{BrownOttewill:1986}
	\begin{align*}
		W(x,x') =w(x) - \tfrac{1}{2} w_{;a}(x)\sigma^{;a} +   \tfrac{1}{2} w_{ab}(x)\sigma^{;a}\sigma^{;b} + \dots
	\end{align*}
	so
	\begin{gather*}
		\left[W(x,x')_{;a'}\right] = \left[W(x,x')_{;a}\right] = \tfrac{1}{2}w_{;a}(x) , \\
		\left[W(x,x')_{;a'b'}\right] = \left[W(x,x')_{;ab}\right]=w_{ab}(x), \\
		\left[W(x,x')_{;a'b}\right] =  \tfrac{1}{2} w_{;ab}(x) - w_{ab}(x),\\ 
		\left[W(x,x')_{;a'b'}\right] = -  \left[W(x,x')_{;a'b}\right] +  [W(x,x')_{;a}]_{;b} \quad,
	\end{gather*}
	where the last line represents a special case of Synge's theorem~\cite{Poisson_2004}.
	
	In addition the wave equation requires~\cite{BrownOttewill:1986}
	\begin{align}
		\label{eq:wave}
		w^{a}{}_{a}  - \xi R w  -\mu^2 w=  -\frac{3}{4\pi^2}v_1
	\end{align}
	where
	\begin{align}
		v_1 &= \tfrac{1}{720}R_{pqrs}R^{pqrs}- \tfrac{1}{720}R_{pq}R^{pq}- \tfrac{1}{24}(\xi-\tfrac{1}{5})\square R\nonumber\\
		&\quad  + \tfrac{1}{8}(\xi-\tfrac{1}{6})^2R^2 + \tfrac{1}{4}\mu^2(\xi-\tfrac{1}{6}) R +\tfrac{1}{8}\mu^4.
	\end{align}

	The Hadamard renormalization prescription then yields (up to the standard renormalization ambiguity) the following definitions~\cite{BrownOttewill:1986}
	\begin{align}
		\langle\hat{\phi}^2\rangle_{\subR} = w(x)  
	\end{align}
	\begin{align}
		\langle \hat{T}_\xi^{a}{}_{b}\rangle_{\subR} & = - w^{a}{}_{b}   -   (\xi-\tfrac{1}{2}) w^{;a}{}_{;b}+ (\xi-\tfrac{1}{4}) \square w \delta^{a}{}_{b}  \nonumber\\
		&\qquad +\xi R^{a}{}_{b} w -  \frac{1}{8\pi^2} v_1 \delta^{a}{}_{b} \\
		&= - w^{a}{}_{b}   -   (\xi-\tfrac{1}{2})\langle\hat{\phi}^2\rangle_{\subR}{}^{;a}{}_{;b}+ (\xi-\tfrac{1}{4}) \square \langle\hat{\phi}^2\rangle_{\subR} \delta^{a}{}_{b}  \nonumber\\
		&\qquad +\xi R^{a}{}_{b} \langle\hat{\phi}^2\rangle_{\subR} -  \frac{1}{8\pi^2} v_1 \delta^{a}{}_{b}.
	\end{align}
	

	In the current context, Eq.~(\ref{eq:wave}) enables us to determine
	\begin{align}
		w^{r}{}_{r}  = -w^{\tau}{}_{\tau}-w^{\theta}{}_{\theta}-w^{\phi}{}_{\phi} -  \xi R w  -\mu^2 w  -\frac{3}{4\pi^2}v_1
	\end{align}
	without requiring any \emph{radial} derivatives of $W$ or correspondingly $G$; we can do all calculations required with our $\Delta r=0$ expressions.  In addition,  our previous work \cite{taylorbreen:2017} allows us to calculate $\langle\hat\phi^2\rangle_{\subR}$ in spherically symmetric space-times with great speed and accuracy so that the required derivatives of $\langle\hat\phi^2\rangle_{\subR}$, which, of course in this case are only functions of $r$, may be easily and accurately obtained by interpolation.
	
	In passing we note that similarly the trace may be written as
	\begin{align*}
		\langle T_\xi^{a}{}_{a}(x)\rangle_{\subR} 
		& =   3( \xi-\tfrac{1}{6} ) \square {\langle\hat\phi^2\rangle_{\subR}}    -\mu^2 {\langle\hat\phi^2\rangle_{\subR}}  + \frac{1}{4\pi^2} v_1 ,
	\end{align*}
	with the conformally invariant case yielding the standard trace anomaly.
	
	One final remark is in order, in general $W$ and correspondingly $w$ and $w_{ab}$ depend on the coupling $\xi$ but in space-times with vanishing Ricci scalar (which, of course, includes electro-vac solutions) they do not, while the stress tensor, which is derived from the functional derivative with respect to $g_{ab}$, does.   However,  our expression above shows that in this case 
	they are all simply related by,  for example, 
	\begin{align*}
		\langle \hat{T}_\xi^{a}{}_{b}\rangle_{\subR} & =  \langle\hat{T}_0^{a}{}_{b}\rangle_{\subR}- \xi\Bigl(  {\langle\hat\phi^2\rangle_{\subR}}^{\!;a}{}_{;b} -\square  {\langle\hat\phi^2\rangle_{\subR}} \delta^{a}{}_{b}   - R^a{}_b \langle\hat\phi^2\rangle_{\subR}\Bigr) .
	\end{align*}
	
	As we have already calculated $\langle\hat\phi^2\rangle_{\subR}$,  all that is required now is to calculate 
	$w^{\tau}{}_{\tau}$ and $w^{\theta}{}_{\theta}=w^{\phi}{}_{\phi}$.	The most straightforward way to do this is to compute coincidence limits of $W(x,x')$ with mixed time derivatives and mixed angular derivatives, i.e.,
	\begin{align}
		&[g^{\tau \tau'}W_{,\tau\tau'}]=-\frac{1}{4\pi^{2}}\sum_{l=0}^{\infty}(2l+1)\sum_{n=1}^{\infty}\frac{n^{2}\kappa^{2}}{f(r)}\mathsf{g}_{nl}(r)\nonumber\\
		&\qquad-\frac{1}{4\pi^{2}}\left\{\mathcal{T}_{10}^{(p)}+\mathcal{D}_{22}^{(-)}+\frac{1}{f(r)}(\mathcal{T}_{11}^{(p)}+\mathcal{D}_{21}^{(-)})\right\},\\
		&[g^{\phi \phi'}W_{,\phi\phi'}]=\frac{1}{16\pi^{2}r^{2}}\sum_{l=0}^{\infty}(2l+1)l(l+1)\sum_{n=0}^{\infty}(2-\delta^{n}_{0})\mathsf{g}_{nl}(r)\nonumber\\
		&\qquad+\frac{1}{4\pi^{2}}\left\{\mathcal{T}_{10}^{(p)}+\mathcal{D}_{22}^{(-)}\right\}.
	\end{align}
	Provided the expansion order $m$ is sufficiently high, the mode sums here converge fast enough to be amenable to calculation on a standard laptop. 
	
	We have therefore adapted the extended coordinate method of Refs.~\cite{taylorbreen:2016,taylorbreen:2017} to the RSET without any significant revisions or generalizations of the method.  We illustrate the utility of the method in the next section where we apply it to compute the RSET for arbitrarily coupled massive scalar fields on Schwarzschild spacetime.
	
	\section{Numerical Implementation in Schwarzschild}
	Implementing the prescription above in Schwarzschild spacetime, while efficient, is still non-trivial but there are a couple of ways that the calculation can be simplified.
	
	The first is in generating the modes themselves. This is by far the most computationally expensive aspect of the calculation. 
	However, one can reduce the amount of modes required by taking a high order expansion of the singular field. Here we choose to take a 6th order expansion (setting $m=6$ in Eq. (\ref{eq:RegParam})) and generate 20 $l$ modes and 10 $n$ modes, which yields the RSET accurate to approximately 10-15 decimal places for the parameter sets considered in this paper. We employed two distinct methods for generating the modes. In the first approach, we generated the modes numerically. The $p_{nl}(r)$ modes were computed by specifying a high-order Frobenius series as the initial value near the horizon and then numerically integrating outwards, while the $q_{nl}(r)$ modes were obtained by assuming a high-order asymptotic expansion near infinity and integrating inwards. In the second approach, the modes were computed without any significant numerical undertaking. The radial equation can be recast in the confluent Heun form \cite{Ronveaux}. This means that $p_{nl}(r)$ and $q_{nl}(r)$ are combinations of confluent Heun functions satisfying appropriate boundary conditions. This is especially advantageous since confluent Heun functions are built in to modern software packages like Mathematica. There is a difficulty, however, that on the Euclidean spacetime, the second linearly independent confluent Heun function is of the ``logarithmic'' type. These are not yet implemented in Mathematica. Nevertheless, one can construct the second linearly independent solution that satisfies the appropriate boundary conditions. Let $\mathsf{H}(q,\alpha,\gamma,\delta,\epsilon;z)$ be the confluent Heun function analytic in the vicinity of $z=0$. Then we have
	\begin{align}
		p_{nl}(r)=z^{n/2}e^{\tilde{\omega}z} \mathsf{H}(q,\alpha,n+1,1,-2\tilde{\omega},-z)
	\end{align}
	where $z=r/(2M)-1$ and 
	\begin{align}
		\tilde{\omega}&=2 M\sqrt{\mu^{2}+n^{2}\kappa^{2}},\qquad  \alpha=\tfrac{1}{4}n^2 +\tilde{\omega}^{2} -(n+2)\tilde{\omega} \nonumber\\
		q&=\tfrac{1}{4}n^{2}-\tfrac{1}{2}n + \tilde{\omega}^{2} + l(l+1)-(n+1)\tilde{\omega}.
	\end{align}
	A second linearly independent solution is
	\begin{align}
		Y_{nl}(r)=z^{n/2}e^{\tilde{\omega}z}\mathsf{H}\Big(q-\alpha,-\alpha,1,n+1,2\tilde{\omega},1+z).
	\end{align}
	We can construct $q_{nl}(r)$ by a linear combination of these solutions
	$ q_{nl}(r)=p_{nl}(r)+\beta_{nl}Y_{nl}(r)$
	where the coefficients $\beta_{nl}$ are chosen so that $q_{nl}(r)$ satisfies the appropriate boundary conditions at infinity. In practice, we can compute the $\beta_{nl}$ as follows. Let $Q_{nl}^{\infty}(r)$ be an asymptotic expansion of $q_{nl}(r)$, then
	\begin{align}
		\label{eq:beta}
		\beta_{nl}\approx\bigl(Q_{nl}^{\infty}(r_{\infty})-p_{nl}(r_{\infty})\bigr)/Y_{nl}(r_{\infty}),
	\end{align}
	evaluated at a large radial distance $r_{\infty}$. This quasi-analytical method gives excellent agreement with the numerical results. We add the caveat that computing the $\beta_{nl}$ requires tremendous working precision in the calculation since the right-hand side of (\ref{eq:beta}) is a quotient of enormously large numbers.
	
	The second practical simplification in our numerical implementation is in the computation of radial derivatives of $\langle \hat{\phi}^{2}\rangle_{\subR}$. Rather than compute the radial derivatives of the individual modes and then performing the mode-sum, we simply construct a high-order interpolation function of $\langle \hat{\phi}^{2}\rangle_{\subR}$ that can be differentiated. The interpolation order was chosen to be sufficiently high to guarantee that derivatives of the vacuum polarization up to second order are smooth.
	
	\section{Results}
	In this section, we present some results of our method applied to the computation of the RSET in the Schwarzschild spacetime. We further discuss the implications for the backreaction and the energy conditions.
	
\begin{figure}
		\begin{subfigure}{.5\textwidth}
			\centering
			\includegraphics[width=.9\linewidth]{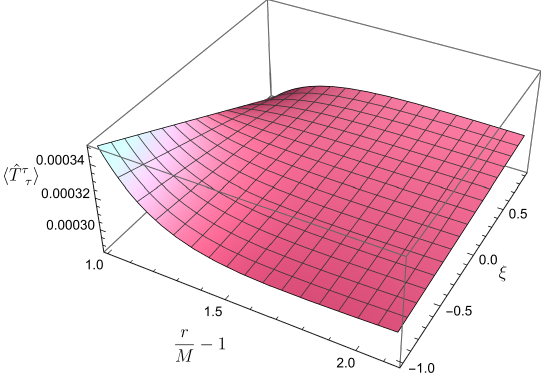}
			\caption{Plot of $\langle \hat{T}^{t}{}_{t}\rangle$ as a function of radius and coupling.}
			\label{fig:Ttt3D}
		\end{subfigure}
		\begin{subfigure}{.5\textwidth}
			\includegraphics[width=.9\linewidth]{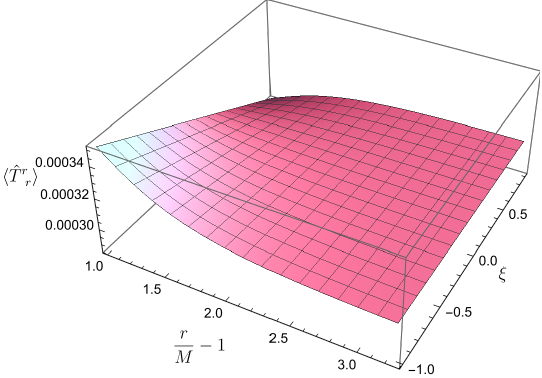}
			\caption{Plot of $\langle \hat{T}^{r}{}_{r}\rangle$ as a function of radius and coupling.}
			\label{fig:Trr3D}
		\end{subfigure}
		\begin{subfigure}{.5\textwidth}
			\vspace{3mm}
			\includegraphics[width=.95\linewidth]{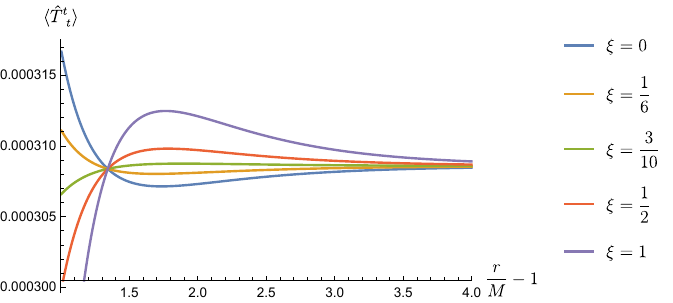}
			\caption{2D plot of $\langle \hat{T}^{t}{}_{t}\rangle$ for various coupling constants.}
			\label{fig:Ttt2D}
		\end{subfigure}
		\caption{Plots showing various components of the RSET and their dependence on the coupling $\xi$.}
		\label{fig:RSETxi}
	\end{figure}

	In Fig.~\ref{fig:RSETxi}, we plot components of the RSET for a fixed field mass $\mu=\frac{1}{2} M$, varying the coupling.  Computing the RSET for massive fields in Schwarzschild is trickier than for massless fields since the Hadamard parametrix contains logarithmic divergences which must be regularized. These are contained in our regularization parameter $\chi_{nl}(r)$. These terms have an arbitrary lengthscale associated with the renormalization ambiguity. We have set this lengthscale to be the mass of the black hole.
	
	We see from Fig.~\ref{fig:RSETxi} that the components of the RSET near the black hole are very sensitive to the coupling $\xi$. For example, for the $\langle T^{t}{}_{t}\rangle_\subR $ components plotted in Fig.~\ref{fig:Ttt2D}, we see that when $\xi$ is approximately less than the conformal value $\xi=1/6$, then this component of the RSET is a decreasing function of $r$ near the horizon, while for $\xi$ approximately greater the conformal coupling, this component of the RSET increases near the horizon. This strong dependence on the coupling presumably implies the backreaction is also sensitive to the coupling.
	
\begin{figure}
		\includegraphics[width=.9\linewidth]{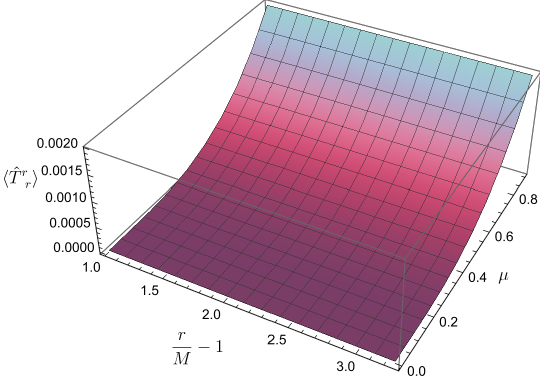}
		\caption{Plot of $\langle \hat{T}^{r}{}_{r}\rangle$ as a function of radius and field mass.}
		\label{fig:Trr3Dmass}
	\end{figure}

	The dependence of the RSET on the mass appears to be a simpler matter. It appears that the components are a monotonically increasing function of the field mass, see Fig.~\ref{fig:Trr3Dmass} for a characteristic plot.  
	
	
\begin{figure}[h]
		\includegraphics[width=.95\linewidth]{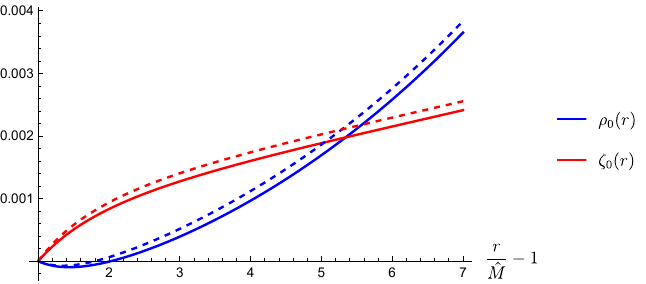}
		\caption{ \label{fig:York} Comparison of the exact solutions to the reduced order field equations with York's approximate results (dashed lines) for a conformal field. Here $\hat{M}$ represents the quantum dressed black hole mass and $\rho_0(r)=\rho(r)-\rho(2\hat{M})$, $\zeta_0(r)=\zeta(r)-\zeta(2\hat{M})$}
	\end{figure}

\begin{figure}[h]
		\includegraphics[width=.95\linewidth]{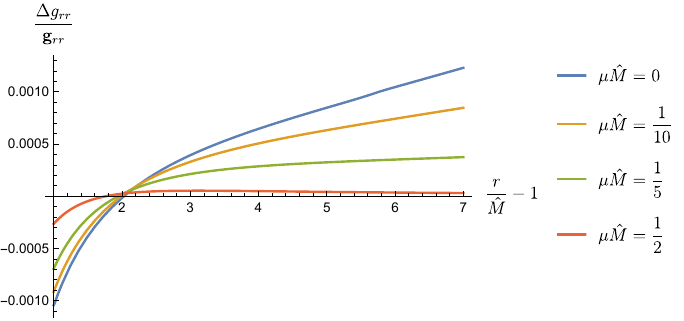}
		\caption{ \label{fig:perturbations}  Ratio of the metric perturbation component $\Delta g_{rr}=2{\hat{M} \zeta_0(r)}/{(r-2\hat{M})}$, to the background metric component $\mathbf{g}_{rr}$, for various field masses with $\xi=1/6$.}
	\end{figure}

	Turning now to the calculation of the backreaction. One can use the exact numerical calculation of the RSET to solve a reduced-order version of the semi-classical equations (see \cite{FlanaganWald1996, taylorbreen:2021} for details of the reduction of order prescription) perturbed about the classical Schwarzschild background. This is achieved by assuming the perturbation respects the symmetry of the background and then expressing the perturbed metric components in the form:
	\begin{align}
		g_{tt}=\Psi^2(r)\left(1-{2\mathcal{M}(r)}/{r}\right),~~g_{rr}=\bigl(1-{2\mathcal{M}(r)}/{r}\bigr)^{-1}.
	\end{align}
	By further expanding about the background metric as $\Psi(r)=1+\hbar \rho(r) + \mathcal{O}(\hbar^2)$ and $\mathcal{M}(r)=M(1+\hbar \zeta(r)) +\mathcal{O}(\hbar^2)$ (reinstating $\hbar$ momentarily for transparency), the reduced-order semi-classical equations give the following simple ODEs for the unknown functions $\rho$ and $\zeta$:
	\begin{align}
		\label{eq:ODEs}
		&\frac{2 M}{r^2} \zeta'(r) -\Lambda= -8 \pi  \langle \hat{T}^{t}_{~t}\rangle_\subR\nonumber\\
		&\frac{2}{r} \rho'(r)=\frac{8 \pi}{1-2M/r}\left(\langle \hat{T}^{r}_{~r} \rangle_\subR-\langle \hat{T}^{t}_{~t} \rangle_\subR\right)
	\end{align}
	where $\Lambda$  corresponds to a renormalisation of the (zero) cosmological constant and is degenerate with the choice of the renormalisation lengthscale in $\langle \hat{T}^{a}_{~b} \rangle_\subR$. Armed with our exact numerical results for the RSET, we may readily solve the above equations for any given set of field parameters. In Fig.~\ref{fig:York}, we investigate the accuracy of York's approximate solution to Eqs. (\ref{eq:ODEs}), obtained via Page's approximation \cite{York1984} for the RSET valid for conformal fields. We find that, as expected, York's solution closely approximates the full, numerical solutions. An issue that arose during York's calculation of the backreaction induced by a conformal field was that the perturbation was unbounded for large $r$ and the backreaction had to be computed in an artificial box and matched to an asymptotically flat solution. However, the backreaction induced by the large-mass approximation to the RSET does not suffer this pathology \cite{taylorbreen:2021, brett:2000}. What about the intermediate field masses? In Fig.~\ref{fig:perturbations}, we plot the ratio of the metric perturbations to the background metric for various values of $\mu$. We see that as $\mu$ increases, the growth of these perturbations decrease and hence the location of the outer boundary where the solution is matched to an asymptotically flat spacetime can be placed further and further away from the black hole. While the boundedness of the backreaction for large $r$ is only strictly true in the $\mu \to \infty$ limit, in practice, provided that $\mu$ is much larger than the black hole temperature, one need not be concerned with the matching procedure employed by York \cite{York1984}, especially since one is usually interested in the backreaction near the black hole.

	Turning, finally, to the application of our method to the investigation of the energy conditions applied to the RSET in the Hartle-Hawking state. In one approach to this investigation, one could try to examine energy conditions by considering the sign of the RSET measured by a local observer, that is, by considering the RSET projected onto a timelike trajectory. However, in general there is a problem with this approach in that the RSET usually depends on the renormalization lengthscale $\ell$ and so it is possible that one could have either sign for the locally measured RSET depending on what choice is made for this lengthscale. Hence, this approach cannot be physically meaningful. 
	
	On the contrary, by a serendipitous cancellation of terms that depend on $\ell$, the energy density along the null circular geodesic at $r=3M$ is independent of the renormalization lengthscale and so is physically meaningful. Applying our method for a range of field masses, we find that, except where the coupling is extremely negative (the semi-classical perturbations are no longer small for such large couplings), the energy density is positive and we find no evidence the Null Energy Condition is violated. This is in contrast to the conclusion for a massless, minimally coupled field in the Unruh state \cite{LeviOri:2016}. Below we give the energy density for some sample field masses:
	\begin{align*}	
		&T^{\mu=0}_{\textrm{null}}\simeq (1.2093\times 10^{-6}+ 6.5390\times 10^{-12} \xi)a^2 \hbar M^{-4}\nonumber\\
		&T^{\mu=1/10}_{\textrm{null}}\simeq (9.1871\times 10^{-7}+6.3792 \times 10^{-12} \xi)a^2 \hbar M^{-4}\nonumber\\
		&T^{\mu=1/5}_{\textrm{null}}\simeq (5.2529 \times 10^{-7}+6.0034 \times 10^{-12} \xi)a^2 \hbar M^{-4}\nonumber\\
		&T^{\mu=1/2}_{\textrm{null}}\simeq (8.5113 \times 10^{-8}+4.6103 \times 10^{-12} \xi)a^2 \hbar M^{-4}\nonumber
	\end{align*}
	where $a$ is arbitrary. 
	\section{Conclusions}	
	In conclusion, we have presented a very efficient mode-sum regularization prescription for computing the RSET for a scalar field in the Hartle-Hawking state on static, spherically symmetric black hole spacetimes. The method offers an ``off-the-shelf'' solution for rapidly computing the RSET without recourse to an expensive numerical undertaking. We show that all of the components of the RSET can be obtained without taking any radial derivatives of the Green function. This is a great simplification as it implies that we can set the radial points together $r=r'$ from the outset of the calculation. We prove the efficiency of the method by computing and presenting results for the RSET in Schwarzschild spacetime for a scalar field with arbitrary mass and coupling. 
	
	Given the efficiency of our method, it is a routine matter to compute the RSET to sufficient accuracy that it can be employed to numerically compute the backreaction. We employ our results for the RSET to compute the semi-classical backreaction on the classical Schwarzschild black hole induced by the stress-energy of a quantum field with various masses. We show that the asymptotic structure of the perturbed spacetime is sensitive to the field mass. The perturbation grows without bound for large $r$ but the rate of this growth is suppressed for more massive fields. Hence the region for which we might expect this semi-classical perturbation to be valid is larger for more massive fields.
	
	As a final application of our results, we investigated the null energy condition on the photon sphere in Schwarzschild. We found no evidence that the NEC is violated except when the coupling constant is very largely negative, but in this case the perturbations to the background are so large that the perturbative expansion on which our semi-classical approximation is based breaks down.
	
	The current work may be extended in several directions. Perhaps the most straightforward extension would be to other quantum states. One could do this either by working on the Lorentzian spacetime from the outset or by leveraging a state subtraction scheme. It would also be of great interest to attempt to extend this regularization scheme to spacetimes with less symmetry assumptions, such as rotating black holes. In Kerr spacetime, there is no Hartle-Hawking state so it would first be necessary to extend the resuts of this paper to other quantum states. Finally, we have presented only results in four spacetime dimensions, it ought to be straightforward to extend these results to arbitrary dimensions.

	%

	\appendix
	\onecolumngrid
	\newpage

	\section*{Appendix: Hadamard Coefficients and Regularization Parameters}
	Below we list the coefficients $\mathcal{D}_{ij}^{(\pm)}(r)$, $\mathcal{T}_{ij}^{(l)}(r)$, $\mathcal{T}_{ij}^{(r)}(r)$ and $\mathcal{T}_{ij}^{(p)}(r)$  to 2nd order. The higher order coefficients can be found in the accompanying Mathematica notebook. The regularization parameters  $\Psi_{ln}^{(\pm)}(i,j|r)$, $\chi_{nl}(i,j|r)$ are listed on the next page.
	\begin{align*}
		&\mathcal{D}_{00}^{(+)}(r)=2,\nonumber\\
		&\mathcal{D}_{10}^{(+)}(r)=-\frac{f(r) \left(r^2 f''(r)-2 r f'(r)+2 f(r)-2\right)}{12 r^2},\nonumber\\
		&\mathcal{D}_{11}^{(+)}(r)=\frac{f(r) \left(r^2 \left(f'(r)^2-4 \kappa ^2\right)-4 f(r) \left(r f'(r)+1\right)+4
			f(r)^2\right)}{24 r^2},\nonumber\\
		&\mathcal{D}_{20}^{(+)}(r)=\frac{1}{2880 r^4}\Bigg(f(r) \Big[-5 r^2 \left(4 \kappa ^2-f'(r)^2\right) \left(r^2 f''(r)-2 r
		f'(r)-2\right)-8 f(r)^2 \left(3 r^3 f^{(3)}(r)-7 r^2 f''(r)+19 r
		f'(r)+10\right)\nonumber\\
		&\qquad\quad+f(r) \left(9 r^4 f''(r)^2-20 r^2 f''(r)+86 r^2 f'(r)^2+4 r f'(r)
		\left(3 r^3 f^{(3)}(r)-14 r^2 f''(r)+20\right)-40 \kappa ^2 r^2+4\right)+76
		f(r)^3\Big]\Bigg)\nonumber\\
		&\mathcal{D}_{21}^{(+)}(r)=-\frac{1}{2880 r^4}\Bigg(f(r) \Big[r^4 \Big(-20 \kappa ^2 f'(r)^2+f'(r)^4+64 \kappa ^4\Big)+r^2 f(r)
		\Big(-20 \kappa ^2 \Big(r^2 f''(r)-6\Big)+120 \kappa ^2 r f'(r)-30 r
		f'(r)^3\nonumber\\
		&\qquad\qquad\qquad\qquad\quad+f'(r)^2 \Big(11 r^2 f''(r)-30\Big)\Big)+4 f(r)^3 \Big(11 r^2 f''(r)-52
		r f'(r)-40\Big)-2 f(r)^2 \Big(10 r^2 f''(r)-67 r^2 f'(r)^2\nonumber\\
		&\qquad\qquad\qquad\qquad\qquad\qquad\qquad\qquad\qquad\qquad\qquad\qquad\qquad\quad +f'(r) \Big(22 r^3
		f''(r)-80 r\Big)+60 \kappa ^2 r^2-28\Big)+104 f(r)^4\Big]\Bigg)\nonumber\\
		&\mathcal{D}_{22}^{(+)}(r)=\frac{f(r)^2 \left(r^2 \left(f'(r)^2-4 \kappa ^2\right)-4 f(r) \left(r f'(r)+1\right)+4
			f(r)^2\right)^2}{1152 r^4},\nonumber\\
		&\mathcal{D}_{11}^{(-)}(r)=-\frac{f'(r)}{6 r}\nonumber\\
		&\mathcal{D}_{21}^{(-)}(r)=\frac{f(r) \left(-9 r f'(r)^2+6 r f(r) \left(r f^{(3)}(r)-2 f''(r)\right)+2 f'(r)
			\left(7 r^2 f''(r)+f(r)+5\right)\right)}{720 r^3}\nonumber\\
		&\mathcal{D}_{22}^{(-)}(r)=\frac{7 r^2 f'(r)^2-10 r f'(r)+r f(r) \left(9 r f''(r)+4 f'(r)\right)-3 f(r)^2+3}{720
			r^4}\nonumber\\
		&\mathcal{T}_{00}^{(l)}(r)=\frac{6 \mu ^2 r^2-(6 \xi -1) \left(r^2 f''(r)+4 r f'(r)+2 f(r)-2\right)}{12 r^2}\nonumber\\
		&\mathcal{T}_{10}^{(l)}(r)=\frac{1}{480 r^4}\Bigg(r^2 f''(r) \left((10 \xi  (3 \xi -1)+1) r^2 f''(r)-10 (6 \xi -1) \left(2 \xi +\mu ^2 r^2\right)\right)+4 (5 \xi  (24 \xi -5)+1) r^2 f'(r)^2\nonumber\\
		&\qquad\qquad+2 r f'(r) \left((5 \xi -1) r^3 f^{(3)}(r)+2 \left(60 \xi ^2-5 \xi -1\right) r^2 f''(r)+40 (1-6 \xi ) \xi
		+10 \mu ^2 (1-12 \xi ) r^2\right)\nonumber\\
		&	\qquad\qquad+2r f(r) \left((80 \xi  (3 \xi -1)+6) f'(r)+r \left((20 \xi  (3 \xi +2)-8) f''(r)+r \left((5 \xi -1) r f^{(4)}(r)+(40 \xi -7) f^{(3)}(r)\right)\right)\right)\nonumber\\ 
		&	\qquad\qquad\qquad\qquad\qquad\qquad\qquad\qquad-40 f(r) \xi  \left(6 \xi +3 \mu ^2	r^2-1\right)+4 (10 \xi  (3 \xi -1)+1) f(r)^2+2 \left(15 \left(2 \xi +\mu ^2 r^2\right)^2-2\right)\Bigg)\nonumber\\
		&\mathcal{T}_{11}^{(l)}(r)=\frac{f(r)}{480 r^4}\Bigg(-2 \left(r^2 f''(r)+2\right) \left((1-5 \xi ) r^2 f''(r)+10 \xi +5 \mu ^2 r^2-2\right)+2 (1-10 \xi ) r^2 f'(r)^2\nonumber\\
		&\qquad\qquad+r f(r) \left((16-80 \xi ) f'(r)+r \left((12-80 \xi )
		f''(r)+r \left(r f^{(4)}(r)+(6-20 \xi ) f^{(3)}(r)\right)+20 \mu ^2\right)\right)+8 (5 \xi -1) f(r)^2\nonumber\\
		&\qquad\qquad\qquad\qquad\qquad\qquad\qquad\qquad\qquad\qquad\qquad\qquad+r f'(r) \left((10 \xi -1) r^3 f^{(3)}(r)+4 (20 \xi -3) r^2 f''(r)+20 (6 \xi -1)\right)\Bigg)\nonumber\\
	\end{align*}
	\begin{align}
		&\mathcal{T}_{10}^{(r)}(r)=\frac{f(r) \left(r^2 \left(f'(r)^2-4 \kappa ^2\right)-4 f(r) \left(r f'(r)+1\right)+4
			f(r)^2\right) \left((1-6 \xi) \left(2-r \left(r f''(r)+4 f'(r)\right)-2 f(r)+2\right)-6 \mu ^2 r^2\right)}{576 r^4}\nonumber\\
		&\mathcal{T}_{10}^{(p)}(r)=\frac{(f(r)-1) \left((1-6 \xi ) \left(2-r \left(r f''(r)+4 f'(r)\right)-2 f(r)\right)-6 \mu ^2 r^2\right)}{144 r^4}\nonumber\\
		&\mathcal{T}_{11}^{(p)}(r)=\frac{f(r) \left(r f'(r)-2 f(r)+2\right) \left((1-6 \xi) \left(2-r \left(r f''(r)+4 f'(r)\right)-2 f(r)\right)-6 \mu ^2 r^2\right)}{144 r^4}.\nonumber\\
	\end{align}
	The regularization parameters $\Psi_{ln}^{(\pm)}(i,j|r)$, $\chi_{nl}(i,j|r)$ are given by
	\begin{align}
		&\Psi_{nl}^{(+)}(i,j|r)=\frac{2^{i-j-1}i!(2i-1)!!(-1)^{n+j}}{\kappa^{2i+2j}r^{2j+2}j!}
		\sum_{p=n-i}^{n+i}\left(\frac{1}{\eta}\frac{\partial}{\partial\eta}\right)^{j}\frac{P_{l}^{-|p|}(\eta)Q_{l}^{|p|}(\eta)}{(i-n+p)!(i+n-p)!}\nonumber\\
		& \Psi_{nl}^{(-)}(i,j|r)=\frac{(1-j)_{2i-j}(-1)^{n+j}}{2\kappa^{2i-2j}r^{2-2j}}\sum_{k=0}^{j}(-1)^{k}\binom{j}{k}\frac{(l+\tfrac{1}{2}+j-2k)}{(l+\tfrac{1}{2}-k)_{j+1}}
		\sum_{p=n-i+j}^{n+i-j}\frac{P_{l+j-2k}^{-|p|}(\eta)Q_{l+j-2k}^{|p|}(\eta)}{(i-j-n+p)!(i-j+n-p)!}\nonumber\\
		&\chi_{nl}(i,j|r)=\begin{cases}
			\displaystyle{	\frac{(-1)^{n}(i-j)!(2j)!}{2\kappa^{2j}r^{2j-2i}}\sum_{k=0}^{1+i-j}(-1)^{k}\binom{1+i-j}{k}\frac{(l+\tfrac{3}{2}+i-j-2k)}{(l+\tfrac{1}{2}-k)_{2+i-j}}	\sum_{p=n-j}^{n+j}\frac{P_{l+i-j+1-2k}^{-|p|}(\eta)Q_{l+i-j+1-2k}^{|p|}(\eta)}{(j-n+p)!(j+n-p)!}} \\ \qquad\qquad\qquad\qquad\qquad\qquad\qquad\qquad\qquad\qquad\qquad\qquad\qquad\qquad\qquad\qquad\qquad\qquad\qquad\textrm{for}\,\,\,\, l>i-j\\ \null \\
			\displaystyle{\frac{2^{i-1}r^{2i-2j}(-1)^{l}}{\pi \kappa^{2j-1}}\left[\frac{d}{d\lambda}(\lambda+1-l)_{l}\left(\frac{2 r^{2}}{\ell^{2}}\right)^{\lambda-i+j}\int_{0}^{2\pi/\kappa}(1-\cos\kappa t)^{j}e^{-in\kappa t}(z^{2}-1)^{(\lambda+1)/2}\mathcal{Q}_{l}^{-\lambda-1}(z)dt\right]_{\lambda=i-j}} \\
			\qquad\qquad\qquad\qquad\qquad\qquad\qquad\qquad\qquad\qquad\qquad\qquad\qquad\qquad\qquad\qquad\qquad\qquad\qquad\textrm{for}\,\,\,\,	l\le i-j.
		\end{cases}\nonumber
	\end{align}
	In the regularization parameters above, we have used the definitions
	\begin{align*}
		\eta=\sqrt{1+\frac{f(r)}{\kappa^{2}r^{2}}},\qquad z=1+\frac{f(r)}{\kappa^{2}r^{2}}(1-\cos\kappa t).
	\end{align*}
	Moreover, $(x)_{n}$ represents the Pochhammer symbol, $P_{\nu}^{\mu}(\eta)$, $Q_{\nu}^{\mu}(\eta)$ represent the associated Legendre functions of the first and second kind, respectively, with the branch cut along the real axis on $(-\infty,1]$. Finally, $\mathcal{Q}_{\nu}^{\mu}(z)$ is Olver's definition of the associated Legendre function of the second kind. In the expression for $\chi_{nl}(i,j|r)$ for $l\le i-j$, it is possible to express this integral in closed form in terms of Euler's beta functions. However, the expression is very cumbersome so we instead express it in integral form. Moreover, this integral is numerically evaluated very rapidly and so the closed form representation is redundant. We also point out that the $\ell$ appearing in $(2r^{2}/\ell^{2})$ is the arbitrary lengthscale associated with the renormalization ambiguity, not to be confused with the quantum mode number $l$ also appearing in this expression.
	
\end{document}